\documentclass[aps,prl,twocolumn,showpacs,superscriptaddress,10pt]{revtex4}

\usepackage{graphicx}
\usepackage{amsmath,amssymb,amstext,amsthm}
\usepackage{bbold}
\usepackage{cancel}
\usepackage{color}
\usepackage{verbatim}
\usepackage{booktabs}
\usepackage[caption=false]{subfig}

\newcommand{\N}{\mathbb{N}}
\newcommand{\Z}{\mathbb{Z}}

\usepackage[utf8x]{inputenc}

\begin{document}

\title{Topological Edge States with Zero Hall Conductivity in
a Dimerized Hofstadter Model}

\author{Alexander Lau}
\affiliation{Institute for Theoretical Solid State Physics, IFW Dresden, 
01171 Dresden, Germany}

\author{Carmine Ortix}
\affiliation{Institute for Theoretical Solid State Physics, IFW Dresden, 
01171 Dresden, Germany}
\affiliation{Institute for Theoretical Physics, Center for Extreme Matter and Emergent Phenomena, Utrecht University, Leuvenlaan 4, 3584 CE Utrecht, Netherlands}

\author{Jeroen van den Brink}
\affiliation{Institute for Theoretical Solid State Physics, IFW Dresden, 
01171 Dresden, Germany}
\affiliation{Department of Physics, TU Dresden, 01062 Dresden, Germany}

\date{\today}

\begin{abstract}

The Hofstadter model is a simple yet powerful Hamiltonian to study quantum Hall physics in a lattice system, manifesting its essential topological states.
Lattice dimerization in the Hofstadter model opens an energy gap at half filling. Here we show that even if the ensuing insulator has a Chern number equal to {\it zero}, concomitantly a doublet of edge states appear that are pinned at specific momenta. 
We demonstrate that these states are topologically protected by inversion symmetry in specific one-dimensional cuts in momentum space, 
define and calculate the corresponding invariants and identify a platform for the experimental detection of these novel topological states.  

\end{abstract}


\maketitle

%%%%%%%%%%%%%%%%%%%%%%%%%%%%%%%%%%%%%%%%%%%%%%%%%%%%%%%%%%%%%%%%%%%%%%%%%%%%%%%%%%%%%%%%%%%%%%%
%%%%%%%%%%%%%%%%%%%%%%%%%%%%%%%%%% INTRODUCTION %%%%%%%%%%%%%%%%%%%%%%%%%%%%%%%%%%%%%%%%%%%%%%%
%%%%%%%%%%%%%%%%%%%%%%%%%%%%%%%%%%%%%%%%%%%%%%%%%%%%%%%%%%%%%%%%%%%%%%%%%%%%%%%%%%%%%%%%%%%%%%%

\emph{Introduction} -- Since the discovery of the quantum Hall effect in 1980~\cite{KDP80}, and its theoretical explanation in terms of the topological properties of the Landau levels~\cite{TKN82,Koh85}, the investigation of topological phases of matter has become
a most active research area.
It has brought forth
the theoretical prediction and experimental 
verification
of a plethora of different topologically nontrivial electronic quantum phases~\cite{BHZ06,KWB07,HXW09,RIR13,PRK15,KaM05,LFY11,LaT13}.
Contrary to their trivial counterparts, topologically nontrivial quantum phases
exhibit 
protected surface
or edge states that are inside the bulk gap.
These topological states are a direct physical
consequence of the topology of the bulk band structure which 
is characterized by a quantized topological invariant~\cite{HaK10,RSF10}.
One of the most celebrated models for the study of topological properties
of matter was introduced by Hofstadter in 1976~\cite{Hof76}. It describes tight-binding electrons
on a rectangular lattice in the presence of a uniform magnetic field and allows the study of the quantum Hall effect on a lattice. Indeed the Hofstadter Hamiltonian harbors the topological chiral edge states that are responsible for the quantized Hall conductivity. 
If the system is perturbed, the quantization stays intact and precise, even if the perturbation introduces additional edge states: any pair of accidentally induced edge states has opposite chirality and therefore yields an exactly zero contribution to the Hall conductivity, which causes the robustness of the quantum Hall effect.

In this Letter we show that in spite of this 
seemingly benign perturbation to the Hofstadter Hamiltonian, a moderate lattice dimerization causes a topological phase transition, spawning counterpropagating edge states not contributing to the Hall conductivity 
that are yet {\it topologically protected}.
We show that the presence of these states can be understood from the topological properties of lower-dimensional cuts of the system, using a mapping of the Hofstadter Hamiltonian on a collection of one-dimensional (1D) Aubry-Andr\'e-Harper (AAH) models~\cite{AuA80,Har55}.  A subset of AAH chains in this collection preserves inversion symmetry, which guarantees the presence of globally topologically protected doublets of end modes to which the edge states are pinned. 
Such end modes are different in nature from the topological edge states found in the context of off-diagonal AAH models~\cite{GSS13}.
To explicitly prove the robustness of the emerging edge states, we define and calculate the topological invariant that protects them, which turns out to be an invariant for inversion-symmetric AAH models. 
Our results thus add a new chapter to the successful history of analogies between Hofstadter and AAH models~\cite{GSS13,KLR12,MCO15,LCC12,KrZ12}. 
Finally we also identify an experimental setup to probe the existence and properties of these new topological edge states.

%%%%%%%%%%%%%%%%%%%%%%%%%%%%%%%%%%%%%%%%%%%%%%%%%%%%%%%%%%%%%%%%%%%%%%%%%%%%%%%%%%%%%%%%%%%%%%%
%%%%%%%%%%%%%%%%%%%%%%%% DIMERIZED HOFSTADTER MODEL %%%%%%%%%%%%%%%%%%%%%%%%%%%%%%%%%%%%%%%%%%%
%%%%%%%%%%%%%%%%%%%%%%%%%%%%%%%%%%%%%%%%%%%%%%%%%%%%%%%%%%%%%%%%%%%%%%%%%%%%%%%%%%%%%%%%%%%%%%%

\begin{figure}[t]\centering
\includegraphics[width=0.99\linewidth] {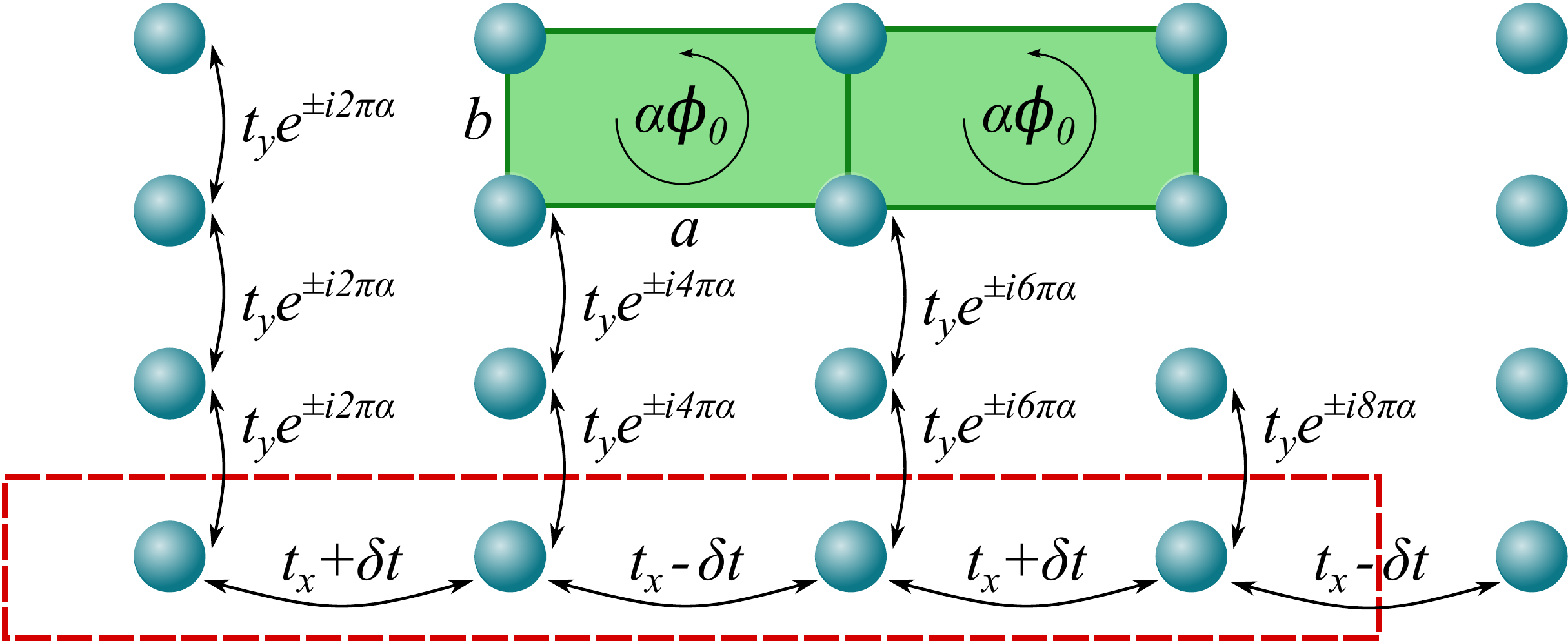}
\caption{(color online). Scheme of the hopping amplitudes in the dimerized
Hofstadter model in the Landau gauge. All plaquettes of size $a\times b$, partially indicated
by green rectangles, are penetrated by a magnetic flux of $\phi=\alpha \phi_0$, 
where $\phi_0=hc/e$ is a magnetic flux quantum. In addition, we show a magnetic
unit cell for the case $\alpha=1/4$ (red dashed rectangle). Note that the unit cell is penetrated
by exactly one flux quantum $\phi_0$.}
\label{fig:dimerized_hofstadter_model}
\end{figure} 

\begin{figure}[t]\centering
\includegraphics[width=0.95\linewidth] {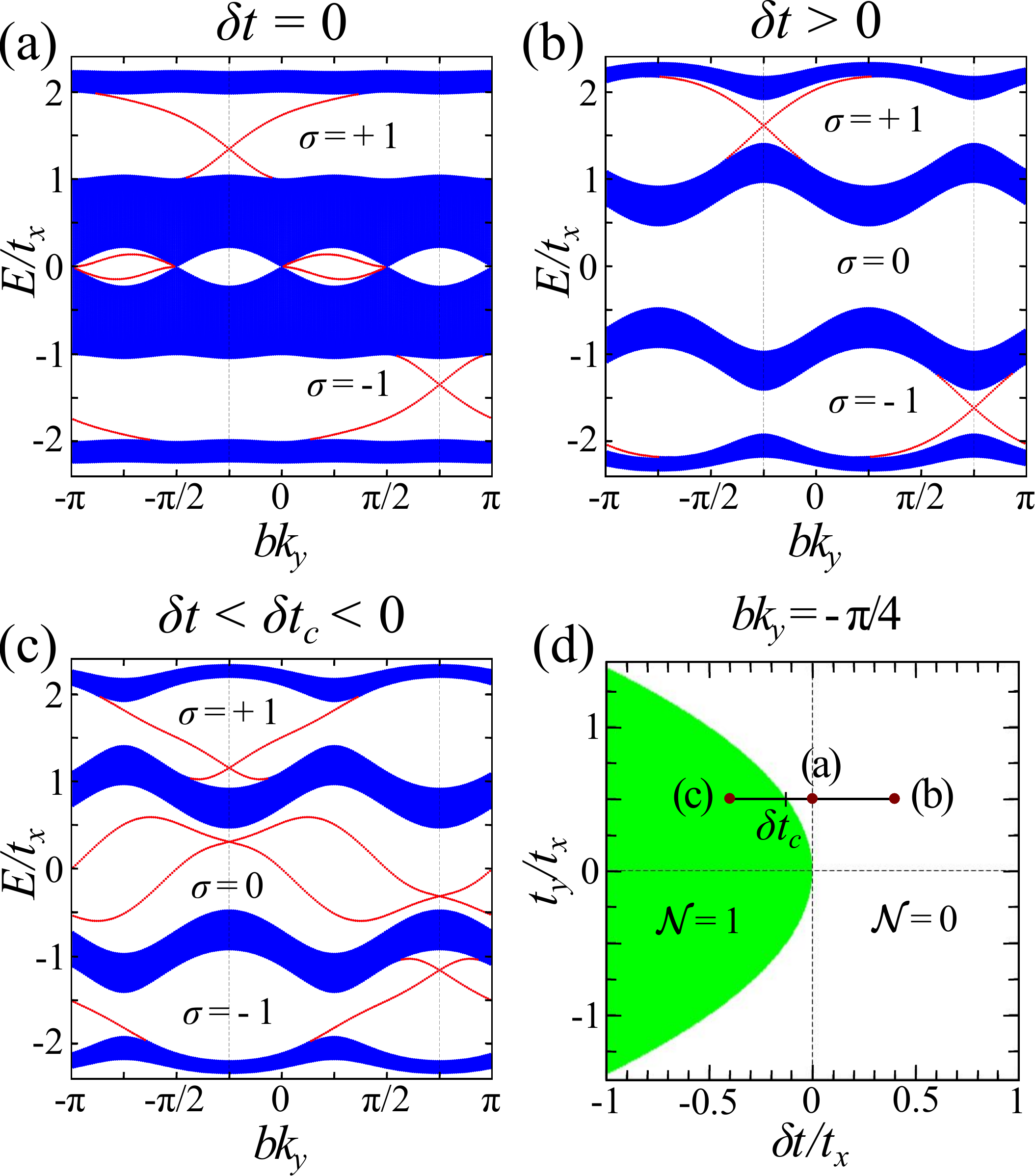}
\caption{(color online). Band structures for the dimerized Hofstadter model in a ribbon 
geometry of width $W=4N_xa$: $\alpha=1/4$, $N_x=100$. Parameters are (in units of $t_x$):  
(a) $t_y=0.5$, $\delta t=0$ (no dimerization).
(b) $t_y=0.5$, $\delta t=0.4$ (trivial dimerization). 
(c) $t_y=0.5$, $\delta t=-0.4$ (nontrivial dimerization). 
States localized at the edges of the system are highlighted in red.
Hall conductivities $\sigma$ for Fermi levels inside the bulk energy gaps are displayed in units
of $e^2/h$. Relevant inversion-symmetric AAH cuts are indicated by dashed vertical lines.
Note that in (c) there are nontrivial edge states at half filling although
the corresponding Hall conductivity is zero. To explain this we show in (d) a half-filling $t_y$-$\delta t$ phase diagram 
for the inversion-symmetric AAH model with $k_yb=-\pi/4$ and with 
respect to the 1D invariant $\mathcal{N}$ of Eq.~\eqref{eq:Z_invariant}. Points corresponding
to (a)--(c) are indicated by red circles, including a possible path connecting them.}
\label{fig:4band_AAH_model_extended_finite}
\end{figure}

\emph{Dimerized Hofstadter model} ---
The Hofstadter model~\cite{Hof76} describes spinless electrons on a rectangular lattice with lattice constants $a$ and $b$ subject to a perpendicular magnetic field. In addition we consider the possibility of a lattice dimerization along one direction being present, which leads to a modulation of hopping amplitudes as indicated in Fig.\ref{fig:dimerized_hofstadter_model}. 
For simplicity, we neglect a modulation of the magnetic fluxes, which would
be present in a realistic system due to the change of the lattice parameter $a$. However, we assure the reader that such a modulation leads to the same general results~\cite{Supp}.
The corresponding tight-binding Hamiltonian, adopting the Landau gauge in a mixed momentum-position space as obtained by performing a Fourier transformation only along the ``undimerized" direction $y$, reads
\begin{eqnarray}
\mathcal{H}&=&\sum_{j_x,k_y} [t_x-(-1)^{j_x}\delta t]\,
(c_{j_x+1,k_y}^\dagger c_{j_x,k_y} + c_{j_x-1,k_y}^\dagger c_{j_x,k_y}) \nonumber\\
&&{}+ \sum_{j_x,k_y} 2t_y\cos(bk_y + 2\pi \alpha j_x)\, c_{j_x,k_y}^\dagger c_{j_x,k_y},
\label{eq:dimerized_Hofstadter_model}
\end{eqnarray}
where $c_{j_x,j_y}^\dagger$, $c_{j_x,j_y}$ are fermionic creation and annihilation operators, $\alpha$ is the magnetic flux in units of the magnetic flux quantum $\phi_0$ penetrating each plaquette of size $a\times b$, $t_{x,y}$ are the nearest-neighbor (average) hopping amplitudes while $\delta t$ parametrizes the dimerization strength.
For simplicity we will restrict ourselves to the case for which $\alpha=1/4$ from here on, but it should be pointed out that the final results are general and hold also for other values of the magnetic flux~\cite{Supp}. 
The magnetic unit cell of the  $\alpha=1/4$ dimerized Hofstadter model contains four inequivalent lattice sites [c.f. Fig.~\ref{fig:dimerized_hofstadter_model}]. 
Therefore, the corresponding Bloch Hamiltonian in full momentum space can be written in terms of the Dirac matrices $\Gamma_i$ and their commutators  $\Gamma_{ij}$ \cite{FuK07}.
For a unit cell going from $j_x=1$ to $j_x=4$, the Hamiltonian reads
\begin{eqnarray}
H &=& 
t_y (\cos bk_y - \sin bk_y)\,\Gamma_5 + t_y (\cos bk_y + \sin bk_y)\,\Gamma_{21}\nonumber\\
&&{}(t_x+\delta t)\,\Gamma_{45} + \frac{1}{2}(t_x-\delta t)(1-\cos 4ak_x)\,\Gamma_{41}\nonumber\\
&&{}+\frac{1}{2}(t_x-\delta t)(1+\cos 4ak_x)\,\Gamma_{23}\nonumber\\
&&{}-\frac{1}{2}(t_x-\delta t)\sin 4ak_x (\Gamma_{24} + \Gamma_{31}). 
\label{eq:4band_Hofstadter_Bloch_Hamiltonian}
\end{eqnarray}
Since the prime physical consequence of nontrivial topological states is the existence of chiral edge states, we study the dimerized Hofstadter model in a  ribbon geometry with periodic boundary conditions in the $y$ direction
and open boundary conditions with a finite number of magnetic unit cells $N_x$ in the $x$ direction.
Thus, the ribbon with $\alpha=1/4$ is of width $W=4N_xa$ and terminated by two boundaries perpendicular
to the dimerization direction. The band
structure of the ribbon is then determined via exact diagonalization of the
first-quantized $4L\times 4L$ Hamiltonian corresponding to Eq.~\eqref{eq:dimerized_Hofstadter_model}. Figure~\ref{fig:4band_AAH_model_extended_finite} shows the ensuing band structure for $t_y=t_x / 2$.

In the absence of dimerization the bulk spectrum is gapped for filling fractions $1/4$ and $3/4$, but gapless
at half-filling with four bulk Dirac points. This is the usual situation of
the Hofstadter spectrum with magnetic flux ratio $\alpha=p/q$ with $p,q\in\N$ and $q=2r$ even~\cite{Hof76,OsA01}:
there are $q-2$ bulk energy gaps whereas the two central bands have
$q$ touching points at $E=0$. Furthermore, within each bulk energy gap we observe 
one pair of counterpropagating edge states traversing the bulk gap 
localized at the edges of the system. Those can be 
attributed to the nontrivial bulk topology for the corresponding filling
levels by bulk-boundary correspondence. Indeed, a calculation of the Chern 
number $n_l$~\cite{TKN82,Koh85} for each energy band $l$ yields Hall conductivities $\sigma=\sum_{l\in\mathrm{occ.}}n_l$ of $\sigma(1/4)=-1$ and $\sigma(3/4)=+1$. This gives rise
to one topologically protected state per edge, consistent with the so-called Diophantine equation~\cite{OsA01}. 

For a finite dimerization mass $\delta t>0$, the modulated hopping amplitude acts as a gap-opening
perturbation at half filling, yielding an additional insulating phase for which we calculate a trivial
Hall conductivity of $\sigma(1/2)=0$. Hence, topologically protected edge states 
are not expected for this phase. Indeed,
the two additional bands of edge states between
the touching points of the two central bands of Fig.~\ref{fig:4band_AAH_model_extended_finite}(a) are pushed into the bulk continuum and 
localized in-gap states are absent
[see Fig.~\ref{fig:4band_AAH_model_extended_finite}(b)]. Moreover, the edge states of the nontrivial insulating states are only slightly deformed signaling that the dimerization mass does not interfere with the bulk topological properties of the system. 

The situation for $\delta t<0$ turns out to be much richer. For small values of  the dimerization mass, 
one again observes the opening of a bulk gap at half filling.
However, at a critical value $\delta t=\delta t_c < 0$ [see Fig.~\ref{fig:4band_AAH_model_extended_finite}(d) for a phase diagram] 
a pair of gap closing and reopening points appears at $bk_y=-\pi/4$ and $bk_y=3\pi/4$. 
Furthermore, in close proximity to these points a pair of counterpropagating chiral edge states [Fig.~\ref{fig:4band_AAH_model_extended_finite}(c)] are revealed.
This is in agreement with the vanishing Hall conductivity since their 
contribution to the Hall current is exactly opposite. By further increasing the dimerization, the corresponding edge bands are deformed.
However, the doublets of in-gap edge states
remain pinned at the momenta $bk_y= -\pi/4$, $3\pi/4$ and cannot be pushed into the bulk {\it independent}  of the value of  $\delta t$ and $t_y$. 

We now show that the presence of such doublets of half-filling in-gap edge states has a topological origin.
In the form of Eq.~\eqref{eq:dimerized_Hofstadter_model}, the dimerized Hofstadter model can be viewed as a collection of dimerized 1D chains parametrized by 
the momentum $k_y$, with periodically modulated on-site potentials of periodicity $1/\alpha$,
amplitude $2t_y$, and phase $bk_y$.  
These chains are equivalent to a specific combination of diagonal and off-diagonal AAH models~\cite{MCO15,LCC12,GSS13,KrZ12}. AAH models have been the subject of intensive research
because of their correspondence to a number of fundamental models, such as 2D lattice models with magnetic flux~\cite{KrZ12},
the Kitaev model~\cite{Kit01}, or the Su-Schrieffer-Heeger model~\cite{SSH79}. Note that the specific form of AAH models studied here is
different from other studies in the literature.
For any value of $k_y$ the AAH models possess time-reversal symmetry with $T=K$, $k_x\rightarrow -k_x$, and with $T^2=+1$,
where $K$ is complex conjugation. 
Moreover, for $\alpha=1/4$ the 1D Hamiltonians preserve inversion symmetry for
four distinct values of $k_y$. In general, one can show that all dimerized Hofstadter models with rational $\alpha=p/q$
possess at least two and at most $2q$ distinct inversion-symmetric cuts~\cite{Supp}.
Note that for a finite number of magnetic unit cells with open boundary conditions 
along the $x$ direction, inversion symmetry persists only in \emph{two} of these chains.
In our example, those cuts are at $k_yb=-\pi/4$ or $3 \pi/4 $, with the 1D parity operator 
described by $P=\sigma^x\otimes\tau^x$ , $\sigma^i$ and $\tau^i$ being
Pauli matrices.

%%%%%%%%%%%%%%%%%%%%%%%%%%%%%%%%%%%%%%%%%%%%%%%%%%%%%%%%%%%%%%%%%%%%%%%%%%%%%%%%%%%%%%%%%%%%%%%
%%%%%%%%%%%%%%%%%%%%%%%% ONE-DIMENSIONAL TOPOLOGICAL INVARIANTS %%%%%%%%%%%%%%%%%%%%%%%%%%%%%%%
%%%%%%%%%%%%%%%%%%%%%%%%%%%%%%%%%%%%%%%%%%%%%%%%%%%%%%%%%%%%%%%%%%%%%%%%%%%%%%%%%%%%%%%%%%%%%%%

\emph{One-dimensional topological invariants} --- The effective 1D inversion-symmetric AAH Hamiltonians fall into the orthogonal class (AI) with inversion symmetry of topological insulators with additional point-group symmetries introduced by Lu and Lee~\cite{LuL14} who thereby extend the famous
Altland-Zirnbauer (AZ) table~\cite{Zir96,AlZ97,HHZ05,SRF08,Kit09,RSF10}.
Note that class AI of the original AZ table is trivial in 1D.
Since inversion operator $P$ ($P^2=+1$) and time-reversal operator $T$ ($T^2=+1$) commute, the 1D Hamiltonians allow for a $\Z$ topological invariant.
Such an integer invariant can be defined as follows~\cite{HPB11}.
Let us consider a 1D system on a chain with inversion symmetry 
described by the Bloch Hamiltonian $H(k)$, $k\in(-\pi/a,\pi/a]$. Inversion symmetry 
implies $P^{-1}H(k)P = H(-k)$ where $P$ is a matrix representation of inversion. In 
particular, $H(k)$ commutes with $P$ at inversion-invariant momenta. Thus,
eigenstates of $H(k)$ have a well-defined parity $\zeta_i(k_\mathrm{inv})=\pm1$ 
at those points. 
The eigenvalues of an operator cannot be changed by continuous deformations
of the Hamiltonian, up to the order. However, a change of the order is only
possible by closing and reopening the energy gap between two states. 
For a 1D inversion-symmetric system, an integer invariant is therefore defined
by\cite{HPB11},
\begin{equation}
\mathcal{N}:= |n_1-n_2|,
\label{eq:Z_invariant}
\end{equation}
where $n_1$ and $n_2$ are the number of negative parities at $k=0$ and $k=\pi/a$,
respectively. 

\begin{figure}[t]\centering
\includegraphics[width=0.95\linewidth] {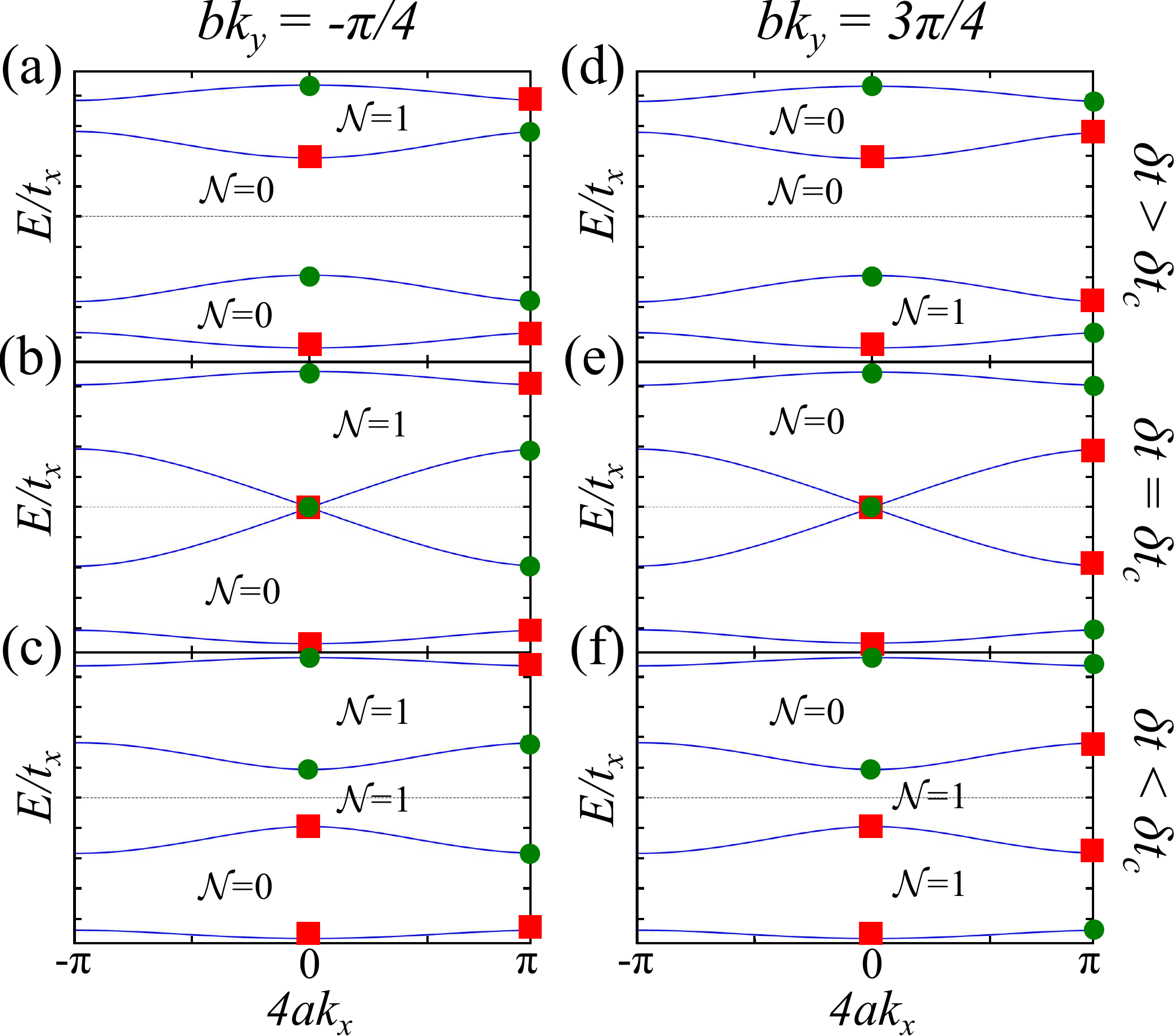}
\caption{(color online). Bulk band structures for inversion-symmetric AAH models with and
without dimerization: $\alpha=1/4$, $t_y=0.5t_x$; $\delta t=0.4t_x$ in (a) and (d), 
$\delta t=\delta t_c=-1/8t_x$ in (b) and (e),
$\delta t=-0.4t_x$ in (c) and (f). The parities at the inversion-invariant momenta
$k_x=0$ and $\pi/4a$ are indicated by green circles ($\zeta=+1$) and red squares
($\zeta=-1$). We also display the topological invariants $\mathcal{N}$ corresponding
to Fermi levels inside the respective bulk energy gaps.}
\label{fig:4band_AAH_model_bulk}
\end{figure}

Let us now apply this to the inversion-symmetric AAH cuts of our exemplary system. In 
Fig.~\ref{fig:4band_AAH_model_bulk}, we show 1D bulk spectra of the 
inversion-symmetric cuts corresponding to Fig.~\ref{fig:4band_AAH_model_extended_finite} for different values of the dimerization mass.
Furthermore, we display the parities of the bulk states at the
inversion-invariant momenta $k_x=0$ and $k_x=\pi/4a$. This enables us to calculate
the topological invariant $\mathcal{N}$. 

As an example, we discuss the results for $bk_y=-\pi/4$ and $t_y>0$
[see Figs.~\ref{fig:4band_AAH_model_bulk}(a)-(c)]. 
Independent of the dimerization mass, we find $\mathcal{N}=1$ for $3/4$ filling 
and $\mathcal{N}=0$ for $1/4$ filling. This means the system is topologically 
nontrivial with one pair of degenerate end modes if the Fermi level is in the upper gap. 
On the contrary, we have a trivial system without end modes for $1/4$ filling.
Going back to the full Hofstadter model, this
explains why the crossing of the $3/4$ filling quantum Hall edge states is pinned to 
the point $bk_y=-\pi/4$.

For half filling, the situation is more subtle. For $\delta t > \delta t_c$, our 
calculations yield $\mathcal{N}=0$, rendering the system
topologically trivial. Indeed, we do not observe end modes in this case. 
In contrast to that, for $\delta t < \delta t_c$ the system is topologically nontrivial
with $\mathcal{N}=1$ and we find a pair of
degenerate end modes in the finite system. The reason for this is that the $k_x=0$ parities
of the two central bands are switched while going from $\delta t > \delta t_c$ to $\delta t < t_c$ -- a band
inversion takes place.

This explicitly explains the observed pinning of the degenerate edge states of the dimerized Hofstadter
model at half filling. They are protected by inversion symmetry in 
specific 1D cuts corresponding to topologically nontrival inversion-symmetric
AAH models. The corresponding
Hall conductivity is zero.

We remark that the qualitative results do not depend on the choice of the unit cell or on 
how the the system is terminated, as long as there is at least one underlying AAH chain
with inversion symmetry. This is always the case for an even number of lattice
sites in the direction of dimerization, whereas for an odd number of sites inversion symmetry
is lost in \emph{all} dimerized AAH chains~\cite{Supp}. However, the details, 
in particular the position of the inversion-symmetric cuts, might change. 

Furthermore, note that the role of the sign of $\delta t$ does depend 
on the termination of the system, which is the usual situation 
for dimerized systems such as the Su-Schrieffer-Heeger model. More specifically, the sign of $\delta t$
is reversed if we consider a dimerized Hofstadter ribbon with unit cells shifted by one lattice site in the
dimerization direction. Physically, this only reflects the fact that the last bonds on both sides of the ribbon
must be weaker than the average bond strength to get nontrivial localized states.

%%%%%%%%%%%%%%%%%%%%%%%%%%%%%%%%%%%%%%%%%%%%%%%%%%%%%%%%%%%%%%%%%%%%%%%%%%%%%%%%%%%%%%%%%%%%%%%
%%%%%%%%%%%%%%%%%%%%%%%%%%%% EXPERIMENTAL DETECTION %%%%%%%%%%%%%%%%%%%%%%%%%%%%%%%%%%%%%%%%%%%
%%%%%%%%%%%%%%%%%%%%%%%%%%%%%%%%%%%%%%%%%%%%%%%%%%%%%%%%%%%%%%%%%%%%%%%%%%%%%%%%%%%%%%%%%%%%%%%

\emph{Experimental detection} -- The Hofstadter model and the AAH model, respectively, have been realized
in different experimental setups such as ultracold atoms in optical lattices~\cite{AAL13,MSK13} 
or photonic crystals~\cite{LPP09,KLR12}. These experimental platforms are characterized by an exceptional 
tunability which brings a much wider range of accessible model parameters into reach. 
A very promising route towards the experimental detection of the pinned degenerate topological states would
be the realization of the inversion-symmetric AAH models.
Similar to the setups described in Refs.~\onlinecite{GSS13} and~\onlinecite{KLR12}, a periodic lattice of 
coupled single-mode waveguides can be prepared on a two-dimensional substrate, where each 
waveguide corresponds to a lattice site of the finite 1D AAH model. The small
spacing between waveguides allows a light wave, propagating through
one of the guides, to tunnel between neighboring waveguides thereby simulating
a hopping process. Furthermore, the width of a waveguide determines
the propagation properties of a light wave, which is used to simulate and 
vary the on-site potentials for different lattice sites. In this way, all
the model parameters of Eq.~\eqref{eq:dimerized_Hofstadter_model} could be implemented
and adjusted. By injecting light
into one of the outermost (boundary) waveguides and by measuring the outgoing
intensity distribution, the localized and topologically protected
end states could then be detected directly.

%%%%%%%%%%%%%%%%%%%%%%%%%%%%%%%%%%%%%%%%%%%%%%%%%%%%%%%%%%%%%%%%%%%%%%%%%%%%%%%%%%%%%%%%%%%%%%%
%%%%%%%%%%%%%%%%%%%%%%%%%%%%%%%% CONCLUSIONS %%%%%%%%%%%%%%%%%%%%%%%%%%%%%%%%%%%%%%%%%%%%%%%%%%
%%%%%%%%%%%%%%%%%%%%%%%%%%%%%%%%%%%%%%%%%%%%%%%%%%%%%%%%%%%%%%%%%%%%%%%%%%%%%%%%%%%%%%%%%%%%%%%

\emph{Conclusions} --- 
We have shown that depending on the sign of the dimerization mass $\delta t$ and on the position of the boundaries,
the dimerized Hofstadter model exhibits topologically
protected edge states at half filling.
The topological states propagate along the edges of a ribbon, perpendicular to the dimerization direction,
whose width must be chosen to extend over an even number of lattice sites.
This has been confirmed by numerical calculations for a model with an exemplary value of the magnetic flux and
generalizes to arbitrary rational values. 
The edge states are protected by 1D inversion symmetry in specific cuts
of the two-dimensional Hofstadter Brillouin zone corresponding to inversion-symmetric AAH models. Most
importantly, they are different from the well-known quantum-Hall edge states because 
their Hall conductivity is zero and they are, thus, protected solely by inversion symmetry.
Moreover, the associated edge bands can be completely disconnected from the bulk continuum of bands.
To uncover the topological nature of the edge states, we have defined and calculated the integer
topological invariant for the 1D inversion-symmetric cuts, which fall in class AI of the
extended classification scheme by Lu and Lee~\cite{LuL14}. These states are thus fundamentally
different from those in the purely off-diagonal AAH model~\cite{GSS13}, which is in the \textit{standard} chiral orthogonal AZ class (BDI).

We have further proposed an experimental setup based on a lattice of coupled
single-mode waveguides that allows for the direct detection of the novel topological
states we predict.

From a more general perspective, we have presented a two-dimensional insulating system where lower-dimensional, 1D physics enriches the global topological structure of the system. Going one step further, it will be very interesting to find realistic materials featuring the ensuing zero-Hall-conductivity topological edge states protected in a reduced dimension.

\emph{Acknowledgements} --- C. O. acknowledges the financial support of the Future and Emerging Technologies (FET) program within the Seventh Framework Programme for Research of the European Commission, under FET-Open Grant No. 618083 (CNTQC), and also the support of the Deutsche Forschungsgemeinschaft (Grant No. OR 404/1-1). We thank SFB 1143 for support.

\clearpage

\newpage

%%%%%%%%%%%%%%%%%%%%%%%%%%%%%%%%%%%%%%%%%%%%%%%%%%%%%%%%%%%%%%%%%%%%%%%%%%%%%%%%%%%%%%%%%%%%%%%
%%%%%%%%%%%%%%%%%%%%%%%%%%%%%%%%% SUPPLEMENTAL MATERIAL %%%%%%%%%%%%%%%%%%%%%%%%%%%%%%%%%%%%%%%
%%%%%%%%%%%%%%%%%%%%%%%%%%%%%%%%%%%%%%%%%%%%%%%%%%%%%%%%%%%%%%%%%%%%%%%%%%%%%%%%%%%%%%%%%%%%%%%

\section*{SUPPLEMENTAL MATERIAL}

\subsection*{A: Inversion-symmetric AAH cuts in the Hofstadter model}

In this section, we are going to show that every dimerized Hofstadter model with rational
magnetic flux $\alpha=p/q$, $p,q\in\Z$ and coprime, has inversion-symmetric cuts for at least
two values of the parameter $bk_y=\in(-\pi,\pi]$.

As discussed in the main paper, the Hofstadter model can be viewed as a collection of
AAH chains parameterized by the momentum $k_y$. For fixed $k_y$, the Hamiltonian 
of such a 1D chain reads,
\begin{eqnarray}
\mathcal{H}_{k_y}&=&\sum_{j_x} [t_x-(-1)^{j_x}\delta t]\,
(c_{j_x+1}^\dagger c_{j_x} + c_{j_x-1}^\dagger c_{j_x}) \nonumber\\
&&{}+ \sum_{j_x} 2t_y\cos(bk_y + 2\pi \alpha j_x)\, c_{j_x}^\dagger c_{j_x}.
\end{eqnarray}
To prove our proposition, we first show that the 1D unit cell can be chosen to be inversion-symmetric.

Let us start with the undimerized case, where the hopping amplitudes $t_x$ are 
constant throughout the lattice. Hence, the kinetic term of the Hamiltonian always preserves inversion symmetry, which
is why we have to focus on the on-site potentials. 
For given $\alpha=p/q$, one period of the on-site potential modulation is $q$ lattice spacings.
Without loss of generality, let us pick a unit cell that goes from $j=1$ to $j=q$.
To have an inversion center in the unit cell, the on-site potentials need to satisfy 
the following condition,
\begin{equation}
\cos\Big(2\pi\,\frac{p}{q}\,j + bk_y \Big) \overset{!}{=} \cos\Big(2\pi\,\frac{p}{q}\,(q-j+1) + bk_y \Big).
\label{eq:inversion_condition}
\end{equation}
This is always true, if we choose $bk_y=-\frac{p\pi}{q}$:
\begin{proof}
\begin{eqnarray}
\mathrm{LHS} &=& \cos\Big(2\pi\,\frac{p}{q}\,j - \frac{p\pi}{q} \Big)
= \cos\Big[\frac{p\pi}{q}\,(2j-1)\Big] \\
\mathrm{RHS} &=& \cos\Big(p - 2\pi\,\frac{p}{q}\,j + 2\pi\,\frac{p}{q} - \frac{p\pi}{q} \Big) \nonumber\\
&=& \cos\Big[\frac{p\pi}{q}\,(1-2j)\Big] =  \cos\Big[\frac{p\pi}{q}\,(2j-1)\Big]\\
&\Longrightarrow& \:\:\mathrm{LHS} = \mathrm{RHS}\nonumber
\end{eqnarray}
\end{proof}
In addition, there is always a second point with $bk_y=-\frac{p\pi}{q}+\pi = \frac{q-p}{q}\pi$ at which
the system obeys inversion symmetry. This case corresponds to substituting $t_y\rightarrow -t_y$.

Let us know check, if this statement also holds for the dimerized Hofstadter model. 
If $q$ is even, a dimerization does not change the size of the magnetic unit cell, which
consists of an even number of sites. The inversion center is exactly in the middle between the two central sites. 
Thus, an alternating hopping amplitude does not violate the inversion symmetry. 

If $q$ is odd, a dimerization doubles the unit cell. Thus, the unit cell
comprises $2q$ lattice sites, which is an even number. Before adding
the dimerization, the inversion center was exactly at the position of the central
lattice site of the undoubled unit cell. Now, it is between the two central sites
of the doubled unit cell. With this, we readily see that if the old unit cell
preserves inversion symmetry, so does the new unit cell in the presence of dimerization. 

As an intermediate result, the values of $k_y$ for which the chosen unit cell
exhibits inversion symmetry are
\begin{equation}
bk_y=-\frac{p\pi}{q} 
\:\:\:\: \mathrm{and} \:\:\:
bk_y= \frac{q-p}{q}\pi.
\end{equation}

However, we have not found all inversion-symmetric cuts, yet. The missing cuts are those
for which the current unit cell itself 
does not have inversion symmetry but the overall system does. To see this, let us now 
choose a unit cell that is shifted by one lattice site. In the onsite
potential term this can be compensated by a phase shift of $2\pi\alpha$.
Incorporating this phase into $bk_y$ and repeating the considerations above,
we find that there are two more inversion-symmetric cuts at $bk_y=-\frac{p\pi}{q} - 2\pi\alpha$
and at $bk_y=\frac{q-p}{q}\pi - 2\pi\alpha$. This procedure can be repeated for all possible
unit cells, i.e., $q$ possibilities for even $q$ and $2q$ possibilites for odd $q$.
Note that each inversion-symmetric cut occurs at least twice. 
Hence, we end up with at most $q$ ($2q$) inversion-symmetric 
AAH chains if $q$ is even (odd).

To summarize, for rational magnetic flux of the form $\alpha=p/q$ with $p,q\in\Z$, 
both the undimerized and the dimerized Hofstadter models have $q$ ($2q$), not necessarily
distinct, inversion-symmetric 
cuts in momentum space. For even $q$, those are at
\begin{equation}
bk_y=\frac{p\pi}{q}\,(1+2n),\:\:\:n=-\frac{q}{2}+1,\ldots,\frac{q}{2},
\end{equation}
whereas for odd $q$, they are at
\begin{equation}
bk_y=\frac{p\pi}{q}\,n,\:\:\:n=-q+1,\ldots,q.
\end{equation}
Note that for $p\neq 1$ the corresponding $bk_y$ values need to be translated back 
to the interval $\in(-\pi,\pi]$.

For chains with a finite number of unit cells and open boundary conditions, inversion
symmetry gets lost in most of these cuts. However, it is still preserved in those 
\emph{two} chains for which these unit cells themselves are inversion-symmetric. 

What is more, inversion symmetry even survives in two of the dimerized chains 
if it comprises an \emph{even} number of sites.
In such a case, the system consists of complete unit cells 
plus an even number $2m$ of additional lattice sites. Let us choose the unit cells such that
there are $m$ of the additional sites at both ends of the chain. We already know that there must be
two $k_y$ values for which the chosen unit cells are inversion-symmetric. From this we immediately see
that our arrangement of unit cells and additional lattice sites is
inversion-symmetric for the same values of $k_y$.

On the contrary, if the chain consists of an \emph{odd} number of sites, the additional
lattice sites cannot be arranged symmetrically around the unit cells. Thus,
inversion symmetry is always broken. Besides, this is already apparent
from the dimerized hopping term in the Hamiltonian, which breaks inversion
symmetry for an odd number of sites.

\subsection*{B: Hofstadter model with $\alpha=1/2$}

The dimerized Hofstadter model with $\alpha=1/2$ is a very interesting special case since it can be mapped
to the famous Su-Schrieffer-Heeger~\cite{SSH79} (SSH) model in the presence of a mass term. More specifically,
in this case the Hofstadter model can be viewed as a collection of SSH models parameterized by $k_y$, 
equivalently to the general correspondence between Hofstadter and AAH model.

The Bloch Hamiltonian is a $2\times 2$ Bloch Hamiltonian,
\begin{eqnarray}
H(k_x,k_y) &=& [(t_x+\delta t) + (t_x-\delta t)\cos (2ak_x)]\,\sigma^x \nonumber\\
&&{}- (t_x-\delta t)\sin (2ak_x)\, \sigma^y \nonumber\\
&&{} - 2t_y\cos (bk_y)\, \sigma^z,
\end{eqnarray}
where $\sigma^{x,y,z}$ are Pauli matrices. In this form, the equivalence to the SSH model is
apparent. The relevant 1D inversion operator is $P=\sigma^x$ with $k_x\rightarrow -k_x$. 
Thus, from the perspective of the SSH model, the system preserves 1D inversion symmetry 
for $bk_y=\pm\pi/2$. 

\begin{figure}[t]\centering
\includegraphics[width=0.95\linewidth] {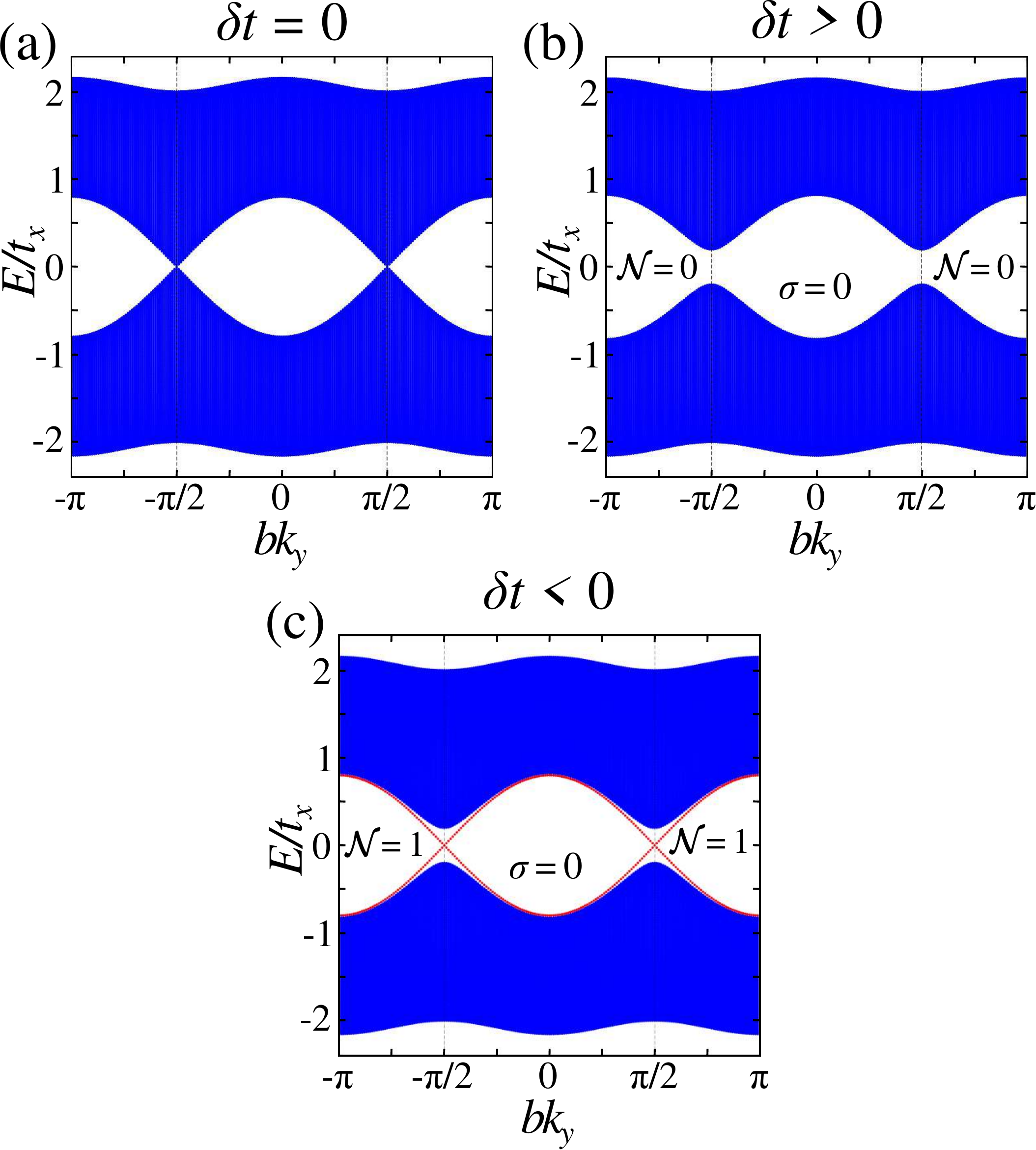}
\caption{(color online). Band structures for the dimerized Hofstadter model in a ribbon 
geometry of width $W=2N_xa$: $\alpha=1/2$, $N_x=200$. Parameters are (in units of $t_x$):
(a) $t_y=0.4$, $\delta t=0$ (no dimerization).
(b) $t_y=0.4$, $\delta t=0.1$ (trivial dimerization). 
(c) $t_y=0.4$, $\delta t=-0.1$ (nontrivial dimerization). 
States localized at the edges of the system are highlighted in red.
Hall conductivities $\sigma$ for Fermi levels inside the bulk energy gaps are displayed in units
of $e^2/h$. The inversion-symmetric AAH cuts are indicated by dashed vertical lines. 
Also, the 1D invariants $\mathcal{N}$ for the inversion-symmetric cuts are shown. Note
that the latter is only nontrivial for $\delta t<0$.}
\label{fig:2band_AAH_model_extended_finite}
\end{figure}

In Fig.~\ref{fig:2band_AAH_model_extended_finite}, we show band structures of the dimerized Hofstadter model
with $\alpha=1/2$ in a ribbon geometry, similar to the one considered in the main paper. 

As expected from general considerations of the Hofstadter model (see main part), in the case without dimerization
the system is gapless with two bulk Dirac points [see Fig.~\ref{fig:2band_AAH_model_extended_finite}(a)].
Furthermore, the Dirac points coincide with the inversion-symmetric cuts of the Hofstadter model. 

For $\delta t>0$, the Dirac points are gapped out without revealing any edge states
[see Fig.~\ref{fig:2band_AAH_model_extended_finite}(b)]. This is readily explained by
a topological analysis: the Hall conductivity at half filling is $\sigma(1/2)=0$, and also the 1D
invariants for the inversion-symmetric cuts at $bk_y=\pm\pi/2$ are trivial with $\mathcal{N}=0$. 
Hence, no topological edge states are expected by bulk-boundary correspondence.

On the contrary, for $\delta t<0$ the topological analysis yields $\mathcal{N}=1$ 
for both inversion-symmetric cuts, giving rise to one pair of
end modes, each, as can be seen in Fig.~\ref{fig:2band_AAH_model_extended_finite}(c). However, the corresponding Hall conductivity
is still zero.

This is in perfect agreement with the results for $\alpha=1/4$ presented in the main part of the paper. It also reflects
what is known for the SSH model~\cite{ETN13}: at the inversion-symmetric cuts, the symmetry-breaking mass term 
associated with $\sigma^z$ vanishes. In this case, the SSH model features topological end modes for a negative
dimerization mass. However, away from those points the SSH model contains a $\sigma^z$ term that gaps out
the end modes. Consequently, those are no longer protected and can now merge with the bulk continuum of bands. 
This is exactly what happens in Fig.~\ref{fig:2band_AAH_model_extended_finite}(c).   

\subsection*{C: Hofstadter model with $\alpha=1/6$}

\begin{figure}[t]\centering
\includegraphics[width=0.95\linewidth] {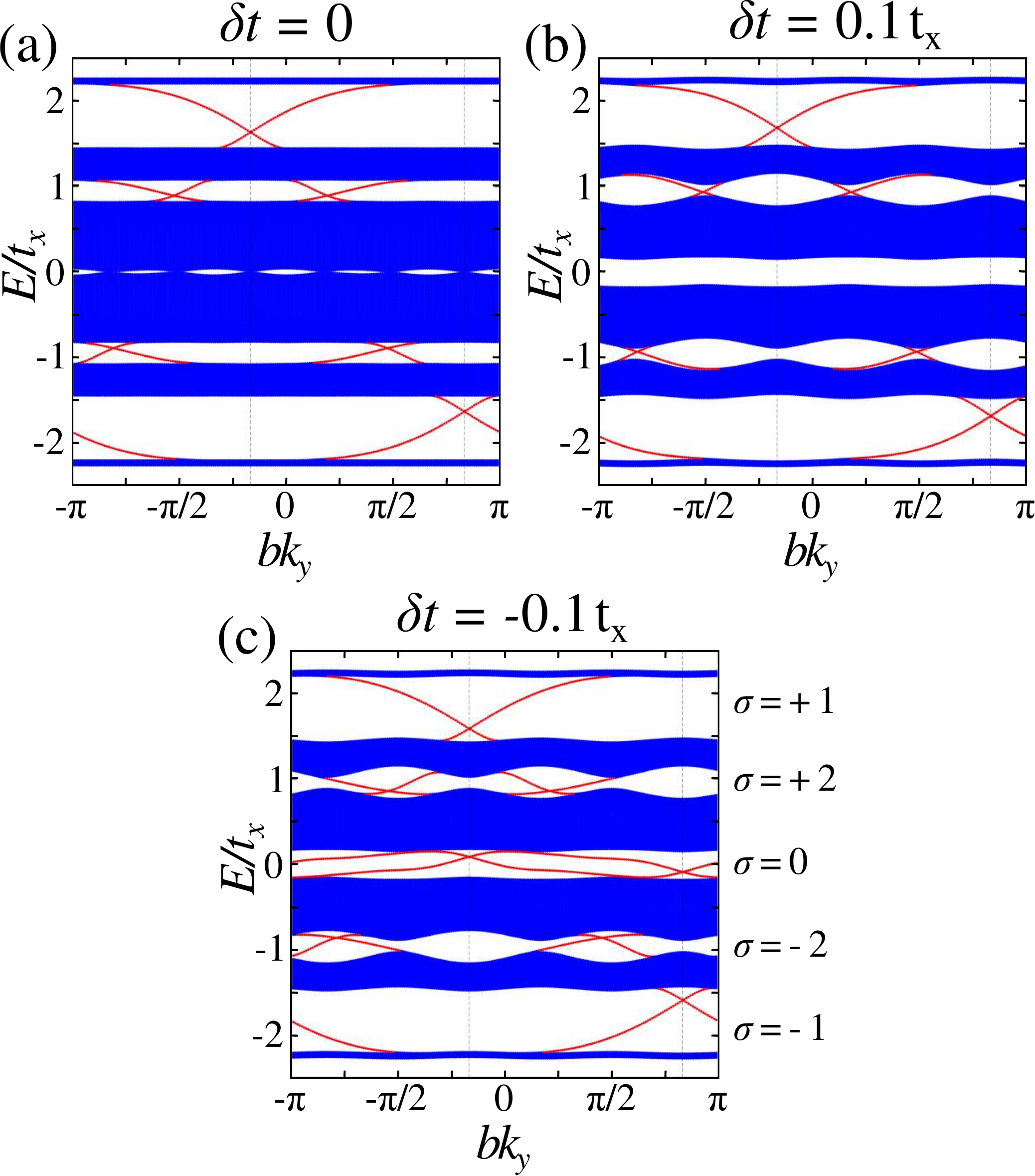}
\caption{(color online). Band structures for the dimerized Hofstadter model in a ribbon 
geometry of width $W=6N_xa$: $\alpha=1/6$, $N_x=70$. Parameters are (in units of $t_x$):
(a) $t_y=0.4$, $\delta t=0$ (no dimerization).
(b) $t_y=0.4$, $\delta t=0.1$ (trivial dimerization). 
(c) $t_y=0.4$, $\delta t=-0.1$ (nontrivial dimerization). 
States localized at the edges of the system are highlighted in red.
Hall conductivities $\sigma$ for Fermi levels inside the bulk energy gaps are displayed in units
of $e^2/h$ for (c) (equivalent gaps in (b) and (c) have the same Hall conductivity).
The relevant inversion-symmetric AAH cuts are indicated by dashed vertical lines.}
\label{fig:6band_AAH_model_extended_finite}
\end{figure}

\begin{figure}[t]\centering
\includegraphics[width=0.95\linewidth] {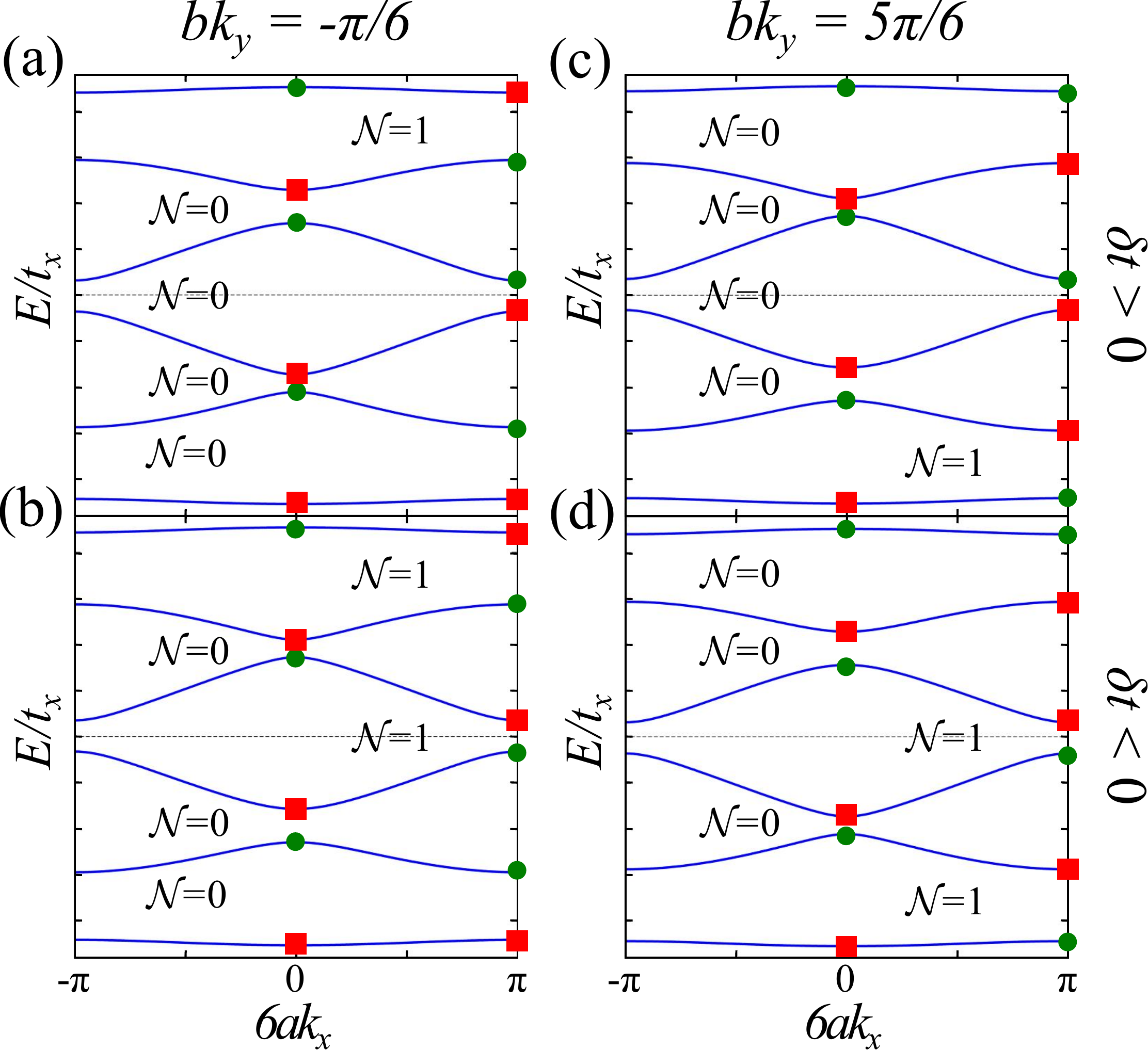}
\caption{(color online). Bulk band structures for inversion-symmetric AAH models with and
without dimerization: $b_V=1/6$, $t_y=0.4t_x$; $\delta t=0.1t_x$ in (a) and (c), 
$\delta t=-0.1t_x$ in (b) and (d). The parities at the inversion-invariant momenta
$k_x=0$ and $\pi/6a$ are indicated by green circles ($\zeta=+1$) and red squares
($\zeta=-1$). We also display the topological invariants $\mathcal{N}$ corresponding
to Fermi levels inside the respective bulk energy gaps.}
\label{fig:6band_AAH_model_bulk}
\end{figure}

In this section, we are going to study the dimerized Hofstadter model
with a magnetic flux of $\alpha=1/6$.
The $6\times 6$ Bloch Hamiltonian $H(k_x,k_y)$ has the following components,
\begin{eqnarray}
H_{nn} &=& 2t_y\cos(\frac{\pi}{3}\,n + bk_y),\:\:\: n=1,\ldots,6 \nonumber\\
H_{n,n+1} &=& H_{n+1,n} = t_x-(-1)^n \delta t,\:\:\: n=1,\ldots,5 \nonumber\\
H_{1,6} &=& H^*_{6,1} = (t_x-\delta t)e^{i6ak_x}. \nonumber
\end{eqnarray}
All other components are zero. The relevant inversion operator $P$ for this representation
is simply the matrix with ones on the anti-diagonal and zeros elsewhere. 

In Fig.~\ref{fig:6band_AAH_model_extended_finite}, we show band structures for this model
in the usual ribbon geometry. The relevant inversion-symmetric cuts are now at $bk_y=-\pi/6$ 
and $bk_y=5\pi/6$ and indicated by vertical dashed lines.
In Fig.~\ref{fig:6band_AAH_model_bulk}, we show the bulk band structures of the corresponding
inversion-symmetric AAH models along with parities and 1D invariants $\mathcal{N}$.

For $\delta t=0$ [see Fig.~\ref{fig:6band_AAH_model_extended_finite}(a)], 
the two central bulk bands touch
each other at six bulk Dirac points. All other bulk bands are separated by a 
finite energy gap. For those gaps, we calculate the Hall conductivities whose
values are in perfect agreement with the number of counterpropagating edge states
we find. As we know from the results of the main part, we can gain
additional topological information from the inversion-symmetric 1D cuts.
At $bk_y=-\pi/6$, only the topmost gap is topologically nontrivial with
$\mathcal{N}=1$, whereas at $bk_y=5\pi/6$ we have the same only for the lowest gap
(compare to Fig.~\ref{fig:6band_AAH_model_extended_finite}, since the topology of these
gaps is not changed for small $\delta t$). Hence, the 2D \textit{and} the 1D topology
of the lowest and topmost gaps are nontrivial and consequently, the crossings
of the topological edge states are pinned to the inversion-symmetric cuts. 

For the second and for the fourth gap, we encounter another interesting situation.
There, we find two pairs of edge states but the crossings are not pinned
to the inversion-symmetric cuts of the BZ. The reason is the trivial 1D topology
with $\mathcal{N}=0$. That means these gaps have a nontrivial 2D topology but no pinned
pairs of edge states.

Let us know study the cases with dimerization. We see that the half-filling gap opens,
but the qualitative results highly depend on the sign of $\delta t$. 
For $\delta t > 0$ [see Fig.~\ref{fig:6band_AAH_model_extended_finite}(b) and Figs.~\ref{fig:6band_AAH_model_bulk}(a),(c)], both the Hall
conductivity and the 1D invariants $\mathcal{N}$ are zero. This is why we do not observe
edge states in this case, as expected by bulk-boundary correspondence.

For $\delta t < 0$ [see Fig.~\ref{fig:6band_AAH_model_extended_finite}(c) and Figs.~\ref{fig:6band_AAH_model_bulk}(b),(d)]
$\sigma$ is also zero, but we have $\mathcal{N}=1$ for both inversion-symmetric cuts. 
Consequently, we find a pair of topological edge states protected by 1D inversion symmetry.

\subsection*{D: Hofstadter model with $\alpha=p/q$ and $q$ odd}

\begin{figure}[t]\centering
\includegraphics[width=0.95\linewidth] {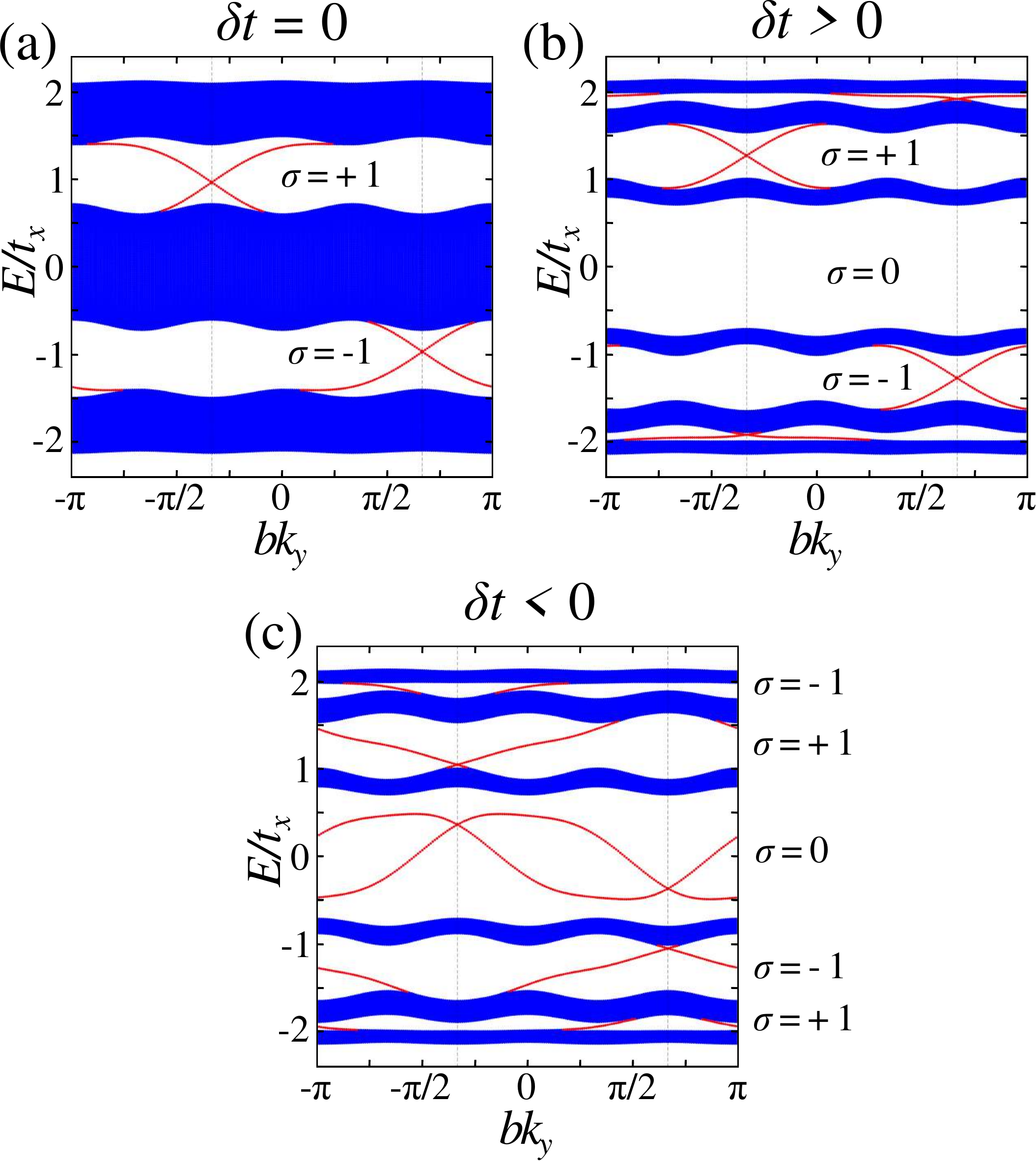}
\caption{(color online). Band structures for the dimerized Hofstadter model in a ribbon 
geometry of width $W=6N_xa$: $\alpha=1/3$, $N_x=60$. Parameters are (in units of $t_x$):
(a) $t_y=0.4$, $\delta t=0$ (no dimerization).
(b) $t_y=0.4$, $\delta t=0.4$ (trivial dimerization). 
(c) $t_y=0.4$, $\delta t=-0.4$ (nontrivial dimerization). 
States localized at the edges of the system are highlighted in red.
Hall conductivities $\sigma$ for Fermi levels inside the bulk energy gaps are displayed in units
of $e^2/h$ for (c) (equivalent gaps in (b) and (c) have the same Hall conductivity).
The relevant inversion-symmetric AAH cuts are indicated by dashed vertical lines.}
\label{fig:3band_AAH_model_extended_finite}
\end{figure}

\begin{figure}[t]\centering
\includegraphics[width=0.95\linewidth] {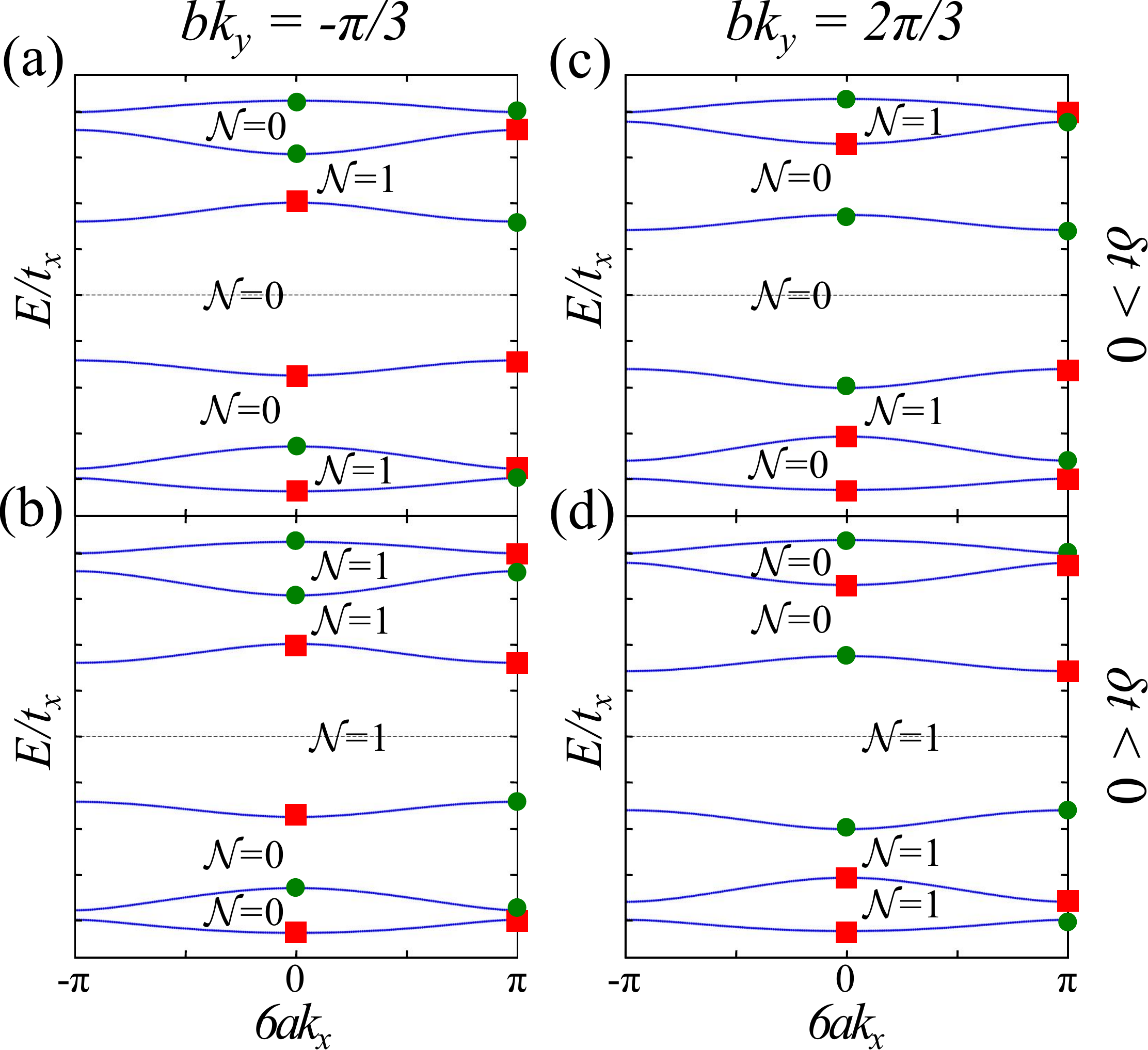}
\caption{(color online). Bulk band structures for inversion-symmetric AAH models with and
without dimerization: $\alpha=1/3$, $t_y=0.4t_x$; $\delta t=0.4t_x$ in (a) and (c), 
$\delta t=-0.4t_x$ in (b) and (d). The parities at the inversion-invariant momenta
$k_x=0$ and $\pi/6a$ are indicated by green circles ($\zeta=+1$) and red squares
($\zeta=-1$). We also display the topological invariants $\mathcal{N}$ corresponding
to Fermi levels inside the respective bulk energy gaps.}
\label{fig:3band_AAH_model_bulk}
\end{figure}

The dimerized Hofstadter model with $\alpha=p/q$ and $q$ odd
needs a separate analysis. This is primarily due to the doubling of the magnetic unit cell
in the presence of dimerization. We will first start with a general discussion supported by an analysis
of numerous Hofstadter models with different values of $\alpha$. After that we will turn
to the special case with $\alpha=1/3$.

In the Hofstadter model, hopping amplitudes w.r.t the $y$ direction are modulated in the $x$ direction 
with a period of $q$ lattice sites, giving
rise to a magnetic unit cell of exactly this size. However, if $q$ is odd, such a period 
is no longer reconcilable with the periodicity of the hopping amplitudes w.r.t the $x$ direction once
a dimerization has been turned on. The greatest common divisor of the two periodicities is $2$,
which is the reason why the magnetic unit cell must be doubled in the dimerized case.

Without dimerization, the bulk band structure of the Hofstadter model with odd $q$ consists of
$q$ bands with $q-1$ full energy gaps separating them. In particular, the Fermi level
of a half-filled system can never be inside one of the gaps. Furthermore, as for the case
with even $q$, a Diophantine equation can be used to infer that the Hall conductivities $\sigma$, associated 
with Fermi levels inside the various gaps, assume every value in the open interval 
$(-\frac{q}{2},\frac{q}{2})$ once and only once~\cite{OsA01}. 

Due to the doubling of the magnetic unit cell in the presence of dimerization, the magnetic
BZ is folded in the direction of $k_x$. This leads to a doubling of the energy bands in the
new magnetic BZ, which now features $2q$ bulk energy bands.
A closer inspection shows that related bands touch along the $k_x=\pm \pi/2qa$ edges 
of the new BZ, if we artificially set the dimerization mass to zero. 
In this case, we have $q$ pairs of bands separated by $q-1$ bulk energy gaps, where the
associated Hall conductivities must be the same as before the folding of the bands.

If we now turn on the dimerization, the touching points within the pairs of bands are gapped
out, revealing another $q$ bulk energy gaps. In total, we thus have $2q-1$ bulk energy
gaps. In particular, there is now also a half-filling gap. We again find
that the half-filling gap is associated with a Hall conductivity of $\sigma=0$.

Independent of the sign of the dimerization mass $\delta t$ the Hall conductivity
vanishes at half filling. Nevertheless, the gap closing at $\delta t=0$ 
switches the parities at $k_x=\pm \pi/2qa$ in the inversion-symmetric cuts, giving
rise to two different topological sectors w.r.t. the 1D invariant $\mathcal{N}$.
Consequently, there are topologically protected edge states for $\delta t<0$ with
zero Hall conductivity, solely protected by inversion symmetry in the 1D cuts. 

Let us now demonstrate the general results for a specific system, namely the
dimerized Hofstadter model with $\alpha=1/3$. The $6\times 6$ Bloch Hamiltonian
$H(k_x,k_y)$ has the following components,
\begin{eqnarray}
H_{nn} &=& 2t_y\cos(\frac{2\pi}{3}\,n + bk_y),\:\:\: n=1,\ldots,6 \nonumber\\
H_{n,n+1} &=& H_{n+1,n} = t_x-(-1)^n \delta t,\:\:\: n=1,\ldots,5 \nonumber\\
H_{1,6} &=& H^*_{6,1} = (t_x-\delta t)e^{i6ak_x}, \nonumber
\end{eqnarray}
while all other components are zero. The relevant 1D inversion operator $P$ for 
this representation is simply the matrix with ones on the anti-diagonal and zeros elsewhere.

In Fig.~\ref{fig:3band_AAH_model_extended_finite}, we show band structures of this 
Hofstadter model in a similar ribbon geometry as in the main part of the paper. To 
explain the 1D topology of the relevant inversion-symmetric cuts, we further plot
1D bulk band structures for the corresponding AAH models in Fig.~\ref{fig:3band_AAH_model_bulk}.

As expected, without dimerization we see two bulk energy gaps with nontrivial Hall
conductivities of $\sigma=\pm 1$, giving rise to the pairs of counterpropagating
edge states we observe in Fig.~\ref{fig:3band_AAH_model_extended_finite}(a). Moreover,
they are pinned to the inversion-symmetric cuts. This is in agreement with the
topological analysis in Fig.~\ref{fig:3band_AAH_model_bulk}. This can be inferred
because the topology of these gaps is independent of $\delta t$.

For nonzero $\delta t$, also the gaps attributed to the folding of the magnetic BZ open. 
Looking at the behavior of the inversion-symmetric cuts in Fig.~\ref{fig:3band_AAH_model_bulk},
we observe that the $k_x=\pi/6a$ parities of three pairs of bands, including the two central
bands, switch by going from $\delta t>0$ to $\delta t<0$. This is a characteristic feature of 
Hofstadter models with odd $q$. For even $q$, this generically happens only for the two central bands.
At half filling, this leads to topological edge states with zero Hall conductivity, but
nontrivial $\mathcal{N}=1$, for $\delta t<0$.

The Hall conductivities at $1/6$ and $5/6$ filling are nontrivial with $\sigma=\pm 1$, independent
of $\delta t$. However, due to the change of the 1D topology at $\delta t=0$ their behavior 
is different for negative and positive $\delta t$. For instance, the crossing of the $1/6$-filling edge states 
is at $bk_y=-\pi/3$ for $\delta t >0$, where it is forced to stay inside the bulk energy gap due to
a nontrivial 1D invariant with $\mathcal{N}=1$. On the contrary, for $\delta t < 0$ the crossing would
be at $bk_y=2\pi/3$, but it has disappeared into the bulk since the 1D invariant $\mathcal{N}$ is trivial.

On the whole, the main features of dimerized Hofstadter models with odd $q$ are the same as for even $q$:
we find topological edge states with zero Hall conductivity at half filling, if the dimerization mass
$\delta t$ is sufficiently negative. They are due to a switch of parities in specific 1D cuts of the 
BZ, giving rise to topological edge states by bulk-boundary correspondence. We merely have to keep in mind
that the width of the ribbon must be compatible with a \textit{doubled} magnetic unit cell.

\subsection*{E: Generalization to arbitrary rational $\alpha=p/q$}

An analysis of numerous Hofstadter models with different rational values
for the magnetic flux $\alpha=p/q$ suggests a generalization of the features
presented in the main part. 
For this, the cases with even and odd $q$ have to be considered separately.

For even $q$, the situation is very similar to the cases $\alpha=1/4$ and $1/2$. 
Without dimerization, the
corresponding Hofstadter models are gapless at half filling because
the two central gaps touch at zero energy. These bulk Dirac points are lifted
by a dimerization mass thereby opening a gap with zero Hall conductivity. 
Furthermore, the system always possesses 
two inversion-symmetric 1D cuts (see Sec.~A). Generically, we find two
situations: either two of the Dirac points coincide with the inversion-symmetric
cuts, or the inversion-symmetric cuts are between two Dirac points. In the
former case, the system exhibits one pair of topological edge states protected
by inversion symmetry for $\delta t<0$, whereas the system is trivial for $\delta t>0$
(c.f. Sec.~B and C with $\alpha=1/2$ and $\alpha=1/6$).
In the other case, for instance for $\alpha=1/4$, the bulk gap closes and reopens 
again at some $t_y$ dependent $\delta t_c<0$. This is the point where the
phase transition from $\mathcal{N}=0$ to $\mathcal{N}=1$ takes place. 
Apart from the shifted transition point, the observations are qualitatively the same 
for both cases. 

For odd $q$, the situation is more subtle. A detailed discussion including the example
$\alpha=1/3$ is given in Sec.~D. First of all, the magnetic unit cell is doubled
due to the dimerization, giving rise to an even number of bands in the folded magnetic
BZ. In this BZ, the nonzero dimerization mass opens a full half filling
gap with zero Hall conductivity, as in the case with even $q$. For $\delta t<0$,
we find again topologically protected edge states pinned to the inversion-symmetric
cuts of the BZ, characterized by a nontrivial 1D invariant $\mathcal{N}$. In contrast
to that, there are no edge states for $\delta t>0$, which can be attributed to both
the zero Hall conductivity and the trivial value for $\mathcal{N}$.

We conclude that the dimerized Hofstadter model with \textit{rational} magnetic flux
$\alpha$ exhibits topologically protected edge states at half filling with zero Hall 
conductivity, if the dimerization mass $\delta t$ is sufficiently small and less than zero. 
The topological states are subject to 1D inversion symmetry in momentum-space cuts
at two specific momenta $k_y$ depending on $\alpha$ and on the way we terminate the system.

\subsection*{F: Modulation of the magnetic flux}

\begin{figure}[t]\centering
\includegraphics[width=0.95\linewidth] {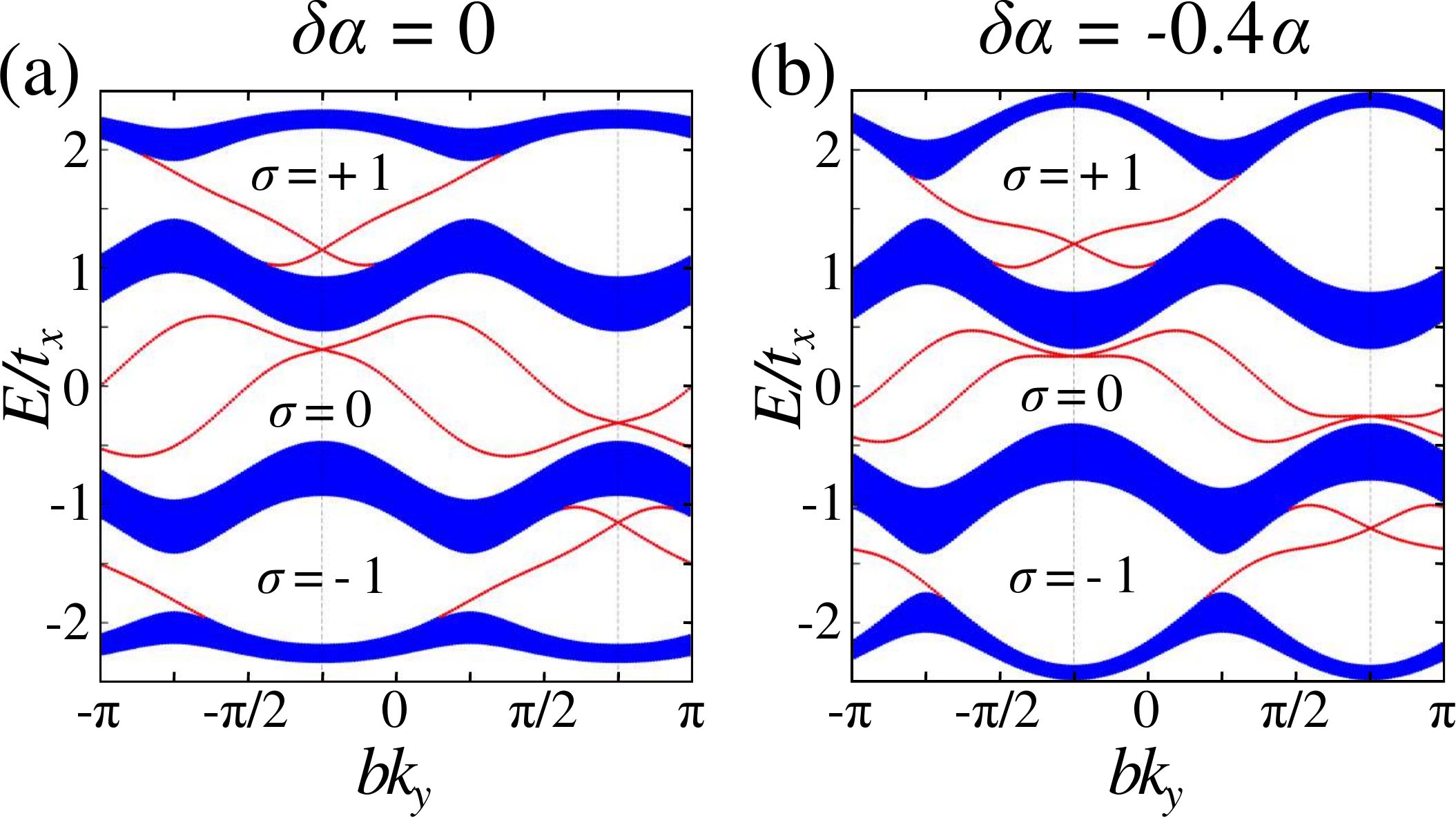}
\caption{(color online). Comparison of the band structures for the dimerized Hofstadter model
with and without flux modulation in a ribbon geometry of width $W=4N_xa$: $\alpha=1/4$, $N_x=100$.
Parameters are (in units of $t_x$):  
(a) $t_y=0.5$, $\delta t=-0.4$, $\delta\alpha=0$
(b) $t_y=0.5$, $\delta t=-0.4$, $\delta\alpha=-0.4\alpha=-0.1$
As usual, states localized at the edges of the system are highlighted in red, and
relevant inversion-symmetric AAH cuts are indicated by dashed vertical lines.
As an example, only the cases with nontrival edge states at half filling are shown.
Note that in (b) the bands are slightly deformed, but the inversion-symmetric
cuts are still at the same values for $k_y$.}
\label{fig:flux_modulation_4band}
\end{figure}

In the following, we are going to discuss how a modulation of the magnetic flux through
adjacent plaquettes of the Hofstadter lattice affects the results of this work.

In a realistic system, a lattice distortion of the kind considered in this Letter
leads to a modulation of the lattice spacing $a\rightarrow a \pm \delta a$ in the
$x$ direction. Since the magnetic flux through each plaquette of the lattice
is proportional to this spacing, also the magnetic flux acquires a modulation, i.e.,
$\phi = \alpha\,\phi_0 \rightarrow \phi \pm \delta\phi = (\alpha\pm\delta\alpha)\phi_0$.
Besides, the hopping amplitudes $t_x$ are modulated as usual. 
Note that the modulations of $t_x$ and $\alpha$ are opposite, i.e., an increase
of $t_x$ corresponds to a decrease of $a$ and thus $\alpha$. Similar to the main 
part of the Letter, this leads us to the
following Hamiltonian in mixed $j_x$-$k_y$ representation,
\vspace{-0.1em}
\begin{eqnarray}
\mathcal{H}&=&\sum_{j_x,k_y} [t_x-(-1)^{j_x}\delta t]\,
(c_{j_x+1,k_y}^\dagger c_{j_x,k_y} + c_{j_x-1,k_y}^\dagger c_{j_x,k_y}) \nonumber\\
&&{}+ \sum_{j_x,k_y} 2t_y\cos[bk_y + 2\pi \alpha j_x - (-1)^{j_x}\pi\,\delta\alpha]\nonumber\\
&&{}\:\:\:\times\,c_{j_x,k_y}^\dagger c_{j_x,k_y},
\label{eq:dimerized_Hofstadter_model}
\end{eqnarray}
We find that the effect of the flux modulation is simply an additional alternating phase
of $\pm\pi\,\delta\alpha$ in the onsite potential term of the corresponding AAH chains.
It is obvious that for rational $\alpha=p/q$ the alternating phase does not change the
translational symmetry of the system as compared to the case without flux modulation.
Furthermore, the condition for inversion symmetry from Sec.~A 
[see Eq.~\eqref{eq:inversion_condition}] changes to
\begin{flalign}
\cos& \Big(2\pi\,\frac{p}{q}\,j + bk_y - (-1)^j\pi\Big) \nonumber\\
&\overset{!}{=}\cos\Big(2\pi\,\frac{p}{q}\,(q-j+1) + bk_y - (-1)^{q-j+1}\pi \Big).
\end{flalign}
It is straight-forward to show that this equation is solved for the same values 
of $k_y$ as in Sec.~A. Hence, the flux modulation does not break inversion
symmetry and in addition, the position of the inversion-symmetric cuts
is not changed. 

In conclusion, this means that the general results of the main part of the Letter
are robust towards a modulation of the magnetic flux.
Merely the details are expected to change, i.e., the energy bands will be slightly
deformed as compared to the case without flux modulation
(see Fig.~\ref{fig:flux_modulation_4band} for $\alpha=1/4$).

\end{document}